\documentclass[11pt,a4paper,twoside]{article}
\usepackage{amssymb,amsfonts,amsmath,amsthm,euscript}
\usepackage{graphicx}
\usepackage{subfigure}
\usepackage{multicol}
\usepackage{multirow}

\theoremstyle{plain}

\newtheorem{prop}{\sc Proposition}[section]
\newtheorem{thm}{\sc Theorem}[section]

\theoremstyle{definition}
\newtheorem{defn}{\sc Definition}[section]
\newtheorem{remark}{\sc Remark}[section]

\newcommand{\cov}{\mathrm{Cov}}
\newcommand{\E}{\mathbb {E}}
\newcommand{\F}[1]{\mathcal{F}_{#1}}
\newcommand{\LL}{\mathcal{B}}
\newcommand{\MaxLoss}{\mathrm{MaxLoss}}
\newcommand{\n}[1]{\mbox{\scriptsize\boldmath{$#1$}}}
\newcommand{\N}{\mathbb {N}}
\newcommand{\nn}[1]{\mbox{\boldmath{$#1$}}}
\newcommand{\noise}{{\small$\{Z_t\}_{t \in \mathbb{Z}}{}$}}
\renewcommand{\P}{\mathbb{P}}
\newcommand{\PP}{\ensuremath{\mathcal{P}}}
\newcommand{\pe}{{\small$\{X_t\}_{t \in \mathbb{Z}}{}$}}

\renewcommand{\proof}{\noindent\textbf{Proof: }}
\newcommand{\R}{\mathbb {R}}

\newcommand{\Z}{\mathbb {Z}}
\newcommand{\ES}[1]{\ensuremath{\mathrm{ES_{p}}{#1}}}
\newcommand{\VaR}[1]{\ensuremath{\mathrm{VaR_{p}}{#1}}}

\begin{document}

%
%

\title{\bf \normalsize RISK MEASURE ESTIMATION ON FIEGARCH PROCESSES\vspace{1.2cm}}

\author{T.S. Prass\footnote{E-mail:
taianeprass@gmail.com} \ and S.R.C. Lopes\footnote{E-mail:
silviarc.lopes@gmail.com}\vspace{0.4cm}\\
Instituto de Matem\'atica - UFRGS\vspace{0.1cm}\\
Porto Alegre - RS - Brazil\vspace{0.4cm}}

\date{This version: November, 2011}

\maketitle

\thispagestyle{empty}

%
%
\noindent
\begin{abstract}
\footnotesize{We consider the Fractionally Integrated
Exponential Generalized Autoregressive Conditional
Heteroskedasticity process, denoted by FIE\-GAR\-CH$(p,d,q)$,
introduced by Bollerslev and Mikkelsen (1996). We present a simulated study regarding  the estimation of the risk measure \VaR{} on FIEGARCH processes. We consider the distribution function of the portfolio log-returns (univariate case) and the multivariate distribution function of the risk-factor changes (multivariate case). We also compare the performance of the risk measures \VaR, \ES{} and MaxLoss for a portfolio composed by stocks of four Brazilian companies.
}
\end{abstract}

\vspace{0.6cm}


\vspace{0.2cm}


%
%

\section{Introduction}\label{introduction}
\renewcommand{\theequation}{\thesection.\arabic{equation}}
\setcounter{equation}{0}

In financial terms, risk is the possibility that an investment will have a return different from the expected, including the possibility of losing part or even all the original investment.  A portfolio is a collection of investments maintained by an institution or a person. In this paper portfolio will be used to indicate a collection of stocks. The selection of an efficient portfolio is an important issue and it is discussed in Prass and Lopes (2009). The authors consider an approach based in the mean-variance (MV) method introduced by Markowitz (1952).

In finance one of the most important problems is risk management, which involves risk measures estimation. Among the applications of a risk measure  we can say that it can be used to determine the capital in risk, that is, we can measure the exposure to the risks of a financial institution, in order to determine the necessary amount  to honour possible unexpected losses. A more detailed study on quantitative risk management, including theoretical concepts and practical examples, can be found in MacNeil et al. (2005).

Financial time series present an important characteristic known as volatility, which can be defined in different ways but it is not directly observable. In financial terms, the volatility is associated to the risk of an asset. The volatility can be seen as the statistic measure of the possibility that the value of an asset will significantly increase or decrease,  several times, in a given period of time. As a risk measure, the volatility can be calculated by different approaches. The most common one, but not unique,  is to use the variance  (or the standard deviation) of the historical rentability of a given investment.

In order to model time series, in the presence of volatility clusters, heteroskedastic models need to  be considered. This class of models consider the variance as a function not only of the time but also of the past observations. The fitted models are then used to estimate the (conditional) mean and variance of the process and these values in turn are used in several risk measures estimation.

Among the most used non-linear models we find the ARCH-type model (Autoregressive Conditional Heteroskedasticity) and its extension. Such models are used to describe the conditional variance of a time series. The ARCH$(p)$ models were introduced by   Engle (1982).  The main assumption of the model is that the random variables \pe{} are not correlated, but the conditional variance depends on the square of the past $p$ values of the process.  This model was generalized by  Bollerslev (1986) with the introduction of the  GARCH$(p,q)$ models (Generalized ARCH). In this model the conditional variance depends not only on the past values of the process but also on the past values of the conditional variance.

ARCH and GARCH models appear frequently in the literature due to their easy implementation. However, this class of models present a drawback. These models do not take into account the signal of the process \pe, since the conditional variance is a quadratic function of those values. In order to deal with this problem,  Nelson (1991) introduced the
EGARCH$(p,q)$ models (Exponential GARCH).  As in the linear case, where the ARFIMA models are presented as a generalization of ARMA models, in the non-linear case the  FIGARCH models (Fractionally Integrated GARCH), introduced by Baillie et al. (1996) and  FIEGARCH models (Fractionally Integrated EGARCH),  introduced by
Bollerslev and Mikkelsen (1996),  appear to generalize the GARCH and EGARCH models, respectively.  A  study on  ARFIMA and FIEGARCH  models, among  other non-linear processes,  can be found in  Lopes (2008) and Lopes and Mendes (2006).

Lopes and Prass (2009) present a theoretical study on FIEGARH process and data analysis. The authors present  several properties of these processes, including  results on their stationarity and their ergodicity. It is shown that the process {\small$\{g(Z_t)\}_{t \in \Z}$},  in the definition of FIEGARCH processes, is a white noise and from this result the authors prove that if {\small$\{X_t\}_{t\in \Z}$} is a FIEGARCH$(p,d,q)$ process then, {\small$\{\ln(\sigma^2_t)\}_{t\in \Z}$} is an ARFIMA$(q,d,p)$ process. Also, under mild conditions, {\small$\{\ln(X_t^2)\}_{t \in \Z}$} is an ARFIMA$(q,d,0)$ process with non-Gaussian innovations. Lopes and Prass (2009) also analyze the autocorrelation and spectral density functions decay of the {\small$\{\ln(\sigma^2_t)\}_{t \in \Z}$} process and the convergence order for the polynomial coefficients that describes the volatility.

The most used methodology to calculate risk measures is based on the assumption that the distribution  of the data is Gaussian. A drawback present in this method is that the distribution function of financial series usually present tails heavier than the normal distribution. A very well known risk measure is the Value-at-Risk (\VaR{)}. Nowadays it has being changed but the most used methods to estimate the \VaR{} consider the Gaussian assumption. Khindanova and Atakhanov (2002) present a comparison study on \VaR{} estimation. The authors consider the Gaussian, empirical and stable distributions. The study  demonstrates that stable modeling captures asymmetry and heavy-tails of returns, and, therefore, provides more accurate estimates of \VaR{}.  It is also known that, even  considering heavier tail distributions, the risk measures are underestimated for events with small occurrence probability  (extreme events).  Embrechts et al. (1997) present ideas on the modeling of extremal events with special emphasis on applications to insurance and finance.

Another very common approach  is to consider the scenario analysis. Usually one does not make any assumption on the   data distribution. Since this analysis does not provide information on an event probability, it should be used as a complementary tool to other risk measure estimation procedures. Even though the scenarios need to be chosen in such a way that they are plausible and for that it is necessary to have an idea of the occurrence probability for each scenario. The maximum loss, denoted by MaxLoss, introduced by Studer (1997), can be viewed either as risk measure or as  a systematic way to perform a stress test (scenario analysis). In a portfolio analysis, this risk measure can be  interpreted  as the worst possible loss in the portfolio value.


This paper is organized as follows. Section \ref{fiegarchsection}
gives some definitions and some properties of
FIEGARCH$(p,d,q)$ processes. Section \ref{riskmeasure} defines some
risk measures considered here and their relationship.
Section \ref{simulation} gives a
simulation study. Section \ref{application} presents a
portfolio analysis.
Section \ref{conclusionsection} concludes the paper.

%
%
\section{FIEGARCH Process}\label{fiegarchsection}
\renewcommand{\theequation}{\thesection.\arabic{equation}}
\setcounter{equation}{0}
\renewcommand{\thethm}{\thesection.\arabic{thm}}
\setcounter{thm}{0}
\renewcommand{\theremark}{\thesection.\arabic{remark}}
\setcounter{remark}{0}

Financial time series present characteristic common to another time series such as, trend, seasonality, outliers and heteroskedasticity.  However, empirical studies show that return (or log-return) series present some stylized facts. We can say that the return series are not i.i.d. although they show little serial correlation, while the series of absolute or squared returns show profound serial correlation, the conditional expected returns are close to zero, volatility appears to vary over time, return series are leptokurtic (or heavy-tailed) and extreme returns appear in clusters. Due to these characteristics, modeling these time series require considering a class of non-linear heteroskedastic models.  In this section we present the \emph{Fractionally
Integrated Exponential Generalized Autoregressive Conditional Heteroskedasticity} process, denoted by FIEGARCH$(p,d,q)$. This class of processes, introduced by
Bollerslev and Mikkelsen (1996), describes the volatility varying on time, volatility clusters (known as  ARCH and GARCH effects), volatility long-range dependence and also asymmetry.


\begin{defn}\label{definitionfiegarch}
Let \pe{} be a stochasric process. Then, \pe{} is a \emph{Fractionally Integrated} EGARCH \emph{process},
denoted by FIEGARCH$(p,d,q)$, if and only if,

\vspace{-.5cm}
{\small\begin{eqnarray}
X_t&=&\sigma_tZ_t,\label{generalproc}\\
\ln(\sigma_t^2)&=&\omega+\frac{\alpha(\LL)}{\beta(\LL)(1-\LL)^d}g(Z_{t-1}),
\ \ \ \ \mbox{\normalsize for all  } t\in\Z,\label{fiegarchprocess}
\end{eqnarray}}

\vspace{-.3cm}
\noindent where $\omega \in \R$, \noise{} is
a process of i.i.d. random variables with zero
mean and variance equal to one, $\alpha(\cdot)$ and $\beta(\cdot)$ are the polynomials
defined by {\small$\alpha(\LL)\equiv\sum_{i=0}^{p}(-\alpha_i)\LL^i$} and {\small$ \beta(\LL)\equiv\sum_{j=0}^{q}(-\beta_j)\LL^j,$} with $\alpha_0=-1=\beta_0$,  $\beta(\LL)\neq 0$, for all $\LL$ such
that $|\LL|\leq 1$. The function $g(\cdot)$ is  defined by

\vspace{-.2cm}
{\small\begin{equation}\label{functiong}
g(Z_{t})=\theta Z_t +
\gamma\big[|Z_t|-\E\big(|Z_t|\big)\big],\hspace{.3cm}  \mbox{for all
} t\in\Z, \ \mbox{and } \theta, \gamma \in \R,
\end{equation}}

\vspace{-.3cm}
\noindent and the operator  {\small$(1-\LL)^d$} is defined as {\small$
(1-\LL)^d = \sum_{k=0}^{\infty}(-1)\,\delta_{d,k}\,\LL^k \equiv \delta_d(\LL),$} where {\small$\delta_{d,0} = -1$} and

\vspace{-.5cm}
{\small\begin{equation*}
\delta_{d,k} = d\times\frac{\Gamma(k-d)}{\Gamma(k+1)\Gamma(1-d)}=
\delta_{d,k-1}\left( \frac{k-1-d}{k}\right), \ \ \ \mbox{\normalsize for all} \  k\geq 1.
\end{equation*}}
\end{defn}
\vspace{.2cm}

\begin{remark}
If $d = 0$, in expression \eqref{fiegarchprocess}, we have an EGARCH$(p,q)$ model.
\end{remark}

Some properties of the {\small$\{g(Z_t)\}_{t \in \Z}$} process can be found in Lopes and Prass (2009). The authors present  a theoretical study on the  FIEGARCH process properties, including results on their stationarity and their ergodicity. The authors also  show that the process {\small$\{g(Z_t)\}_{t \in \Z}$} is a white noise and use this result to prove that if {\small$\{X_t\}_{t\in \Z}$} is a FIEGARCH$(p,d,q)$ process then, {\small$\{\ln(\sigma^2_t)\}_{t\in \Z}$} is an ARFIMA$(q,d,p)$ process. Moreover, under mild conditions, {\small$\{\ln(X_t^2)\}_{t \in \Z}$} is an ARFIMA$(q,d,0)$ process with non-Gaussian innovations. The autocorrelation and spectral density functions decay of the {\small$\{\ln(\sigma^2_t)\}_{t \in \Z}$} process and the convergence order for the polynomial coefficients that describes the volatility are also analyzed in Lopes and Prass (2009).

In the literature one can find different definitions of FIEGARCH processes. Definition \ref{definitionfiegarch} is the same as in Bollerslev and Mikkelsen (1996). In Zivot and Wang (2005), expression \eqref{fiegarchprocess} is replaced by expression \eqref{fiegarch3} in the definition of a FIEGARCH process.
The following proposition shows that, under the  restrictions  given in \eqref{coeffequiv},
expressions \eqref{fiegarch3} and \eqref{fiegarchprocess} are equivalent. This result is
crucial for a Monte Carlo simulation study (see Section \ref{simulation}).


\begin{prop}\label{equiv}
Let \pe{}  be a \emph{FIEGARCH$(p,d,q)$} process, given in Definition
\ref{definitionfiegarch}.  Then, the expression  \eqref{fiegarchprocess}  can be rewritten as

\vspace{-.55cm}
{\small\begin{eqnarray}
\beta(\LL)(1-\LL)^d\ln(\sigma_t^2)&=& a +
\sum_{i=0}^{p}\Big(\psi_i|Z_{t-1-i}|+\gamma_i Z_{t-1-i}
\Big),\label{fiegarch3}
\end{eqnarray}}
\vspace{-.25cm}
\small{
\begin{equation}
\hspace{-.01cm}\mbox{where} \quad a\equiv (-\gamma)\,\alpha(1)\E\big(|Z_{t}|\big), \quad \psi_i = -\gamma\alpha_{i} \quad  \mbox{ and } \quad
\gamma_i = -\theta\alpha_{i}, \quad \mbox{ for all } \ 0\leq i \leq p.\quad \quad \label{coeffequiv}
\end{equation}
}
\end{prop}

\proof  See Lopes and Prass (2009).\qed

\vspace{.2cm}
Clearly the definition given by Zivot and
Wang (2005) is more general than the one presented by  Bollerslev and
Mikkelsen (1996). In fact, in the definition given by Zivot and Wang (2005), the coefficients  $\psi_j$ and $\gamma_j$, for $j=0,1, \cdots,p$, do not necessarily satisfy the restrictions given in \eqref{coeffequiv}.

%
%

\section{Risk Measures}\label{riskmeasure}
\renewcommand{\theequation}{\thesection.\arabic{equation}}
\setcounter{equation}{0}

In this section we present the concept of risk factor, loss distribution and the definition of some risk measures and some approaches for estimating them. We also show that different approaches can lead to equivalent results depending on the assumptions made.

Risk measures are directly related to risk management. McNeil et al. (2005) classify the existing approaches to measuring the risk of a financial position in four different groups: \emph{Notional-amount approach, Factor-sensitivity measures, Risk measures based on loss distribution and Scenario-based risk measures} (see McNeil et al., 2005, page 34). A frequent concept associated to risk is the volatility, which can have different definitions. The most common approach is to define the volatility as the variance (or the conditional variance) of the processes. In this paper we focus our attention to the variance, the  Value-at-Risk (\VaR) and  Expected Shortfall (\ES), which are risk measures based on loss distributions, and the Maximum-loss, a scenario-based risk measure.

\subsubsection*{Risk Factors}
Consider a given portfolio and denote its value at time $t$ by {\small$V(t)$} (we assume that {\small$V(t)$} is  known at  time $t$). For a given time horizon $h$ we denote by {\small$L_{t+h}$} the loss of the portfolio in the period {\small$[t,t+h]$}, that is, {\small$L_{t+h} := -(V(t+h) - V(t))$}.  The distribution of the random variables {\small$L_{t+h}$} is termed \emph{loss distribution}. In risk management, the main concern is to analyze the probability of large losses, that is, the right tail of the loss distribution.

The usual approach is to assume that the random variable {\small$V_t$}, for all $t$,  is a function of time and an $m$-dimensional random vector {\small$\nn Z_t=(Z_{1,t},\cdots,Z_{m,t})'$} of \emph{risk factors}, that is, {\small$V_t = f(t, \nn Z_t)$}, for some measurable function  {\small$f:\R\times\R^m \rightarrow \R$}. Since we assume that the risk factors are observable,
{\small$\nn Z_t$} is known at time $t$. The choice of  the risk factors and of the function {\small$f(\cdot)$} is a modeling issue and depends on the portfolio and on the desired level of accuracy.

\begin{remark}
The random vector {\small$\nn Z=(Z_1,\cdots,Z_m)'$} is also called a \emph{scenario} which describes the situation of the market and, consequently, {\small$f(\nn Z)$} is referred to as the value of the portfolio under the scenario {\small$\nn Z$}.
\end{remark}

 In some cases it is more convenient to consider instead the time series of \emph{risk-factors change} {\small$\{\nn X_t\}_{t\in\N}$}. This time series is defined by {\small$\nn X_t \equiv  \nn Z_t - \nn Z_{t-1}$}, for all $t \in \N$. For instance, if we consider a portfolio with $m$ stocks and  {\small$ Z_{i,t} = \ln(P_{i,t})$}, where {\small$P_{i,t}$}  is the price of the $i$-th asset at time $t$, then {\small$\nn X_t=(X_{1,t},\cdots,X_{m,t})'$} is the vector of log-returns.  In this case, for $h=1$, the value of the portfolio can be written as {\small$ L_{t+1} = -\big(f(t+1,\nn Z_t+\nn X_{t+1}) - f(t, \nn Z_t)\big)$}, where {\small$\nn Z_t$} is known at time $t$. The loss distribution is then determined by  {\small$\nn X_{t+1}$} risk-factor change distribution.

\subsection{Risk Measures Based on Loss Distributions}\label{subvar}
The variance of the loss distribution is one of the most used risk measures. However, as a risk measure, the variance presents two problems. First of all one needs to assume that the loss distribution has finite second moment. Moreover, this measure does not distinguish between positive and negative deviations of the mean. Therefore, the variance is a good risk measure only for distributions that are (approximately) symmetric, such as the Gaussian or t-Student (with finite variance) distributions.

Another common approach is the quantile analysis of the loss distribution. Consider a portfolio $\PP$ with some risky assets and a fixed time horizon $h$. Let  {\small$F_L(\ell)\equiv \P(L\leq \ell)$} be the distribution of the associated loss. The main idea is to define a statistic based on  {\small$F_L(\cdot)$} capable of measuring the risk of the portfolio over a period $h$. Since for several models the support of {\small$F_L(\cdot)$} is unbounded, the natural candidate (which is the maximum possible loss), given by {\small$ \inf\big\{\ell \in \R: F_L(\ell) =1\big\}$}, is not the best choice. The idea is to consider instead the ``maximum loss which is not exceeded with a high probability''. This probability is called \emph{confidence level}.

\begin{defn}
Let $\PP$ be a fixed portfolio. Given a confidence level  p $\in(0,1)$, the  \emph{Value-at-Risk} of the portfolio, denoted by {\small$\mathrm{VaR}_{\mathrm{p}}$}, is defined as

\vspace{-.3cm}
{\small\begin{equation}
\mathrm{VaR}_{\mathrm{p}} \equiv \inf\{\ell \in \R : \P(L\geq \ell)\leq 1-\mathrm{p}\}= \inf\{\ell \in \R : F_L(\ell)\geq \mathrm{p}\}.
\end{equation}}
\end{defn}

In a probabilistic sense,  $\mathrm{VaR}_{\mathrm{p}}$ is the p-quantile of the loss distribution function. For practical purposes, the most commonly used values are   p $\in\{0.95,0.99\}$ and $h \in\{1,10\}$ days.

As mentioned before,  the risk analysis by considering the variance presents some drawbacks and so does the  {\small$\mathrm{VaR}_{\mathrm{p}}$}.  Artzner \emph{et al.} (1999) define coherent risk measures and show that the {\small$\mathrm{VaR}_{\mathrm{p}}$}  does not satisfy the subadditivity axiom. That is, given a fixed number of portfolios, the {\small$\mathrm{VaR}_{\mathrm{p}}$} of the sum of the portfolios may not be bounded by the sum of the  {\small$\mathrm{VaR}_{\mathrm{p}}$} of the individual portfolios. This result contradicts the idea that the risks can be decreased by diversification, that is, by buying or selling financial assets.

In the following definition we present a coherent risk measure in the sense of Artzner \emph{et al.} (1999) definition.

\begin{defn}\label{es}
Let  $L$ be a loss with distribution function {\small$F_L(\cdot)$}, such that {\small$\E(|L|)<\infty$}. The \emph{Expected Shortfall}, denoted by {\small$\mathrm{ES}_{\mathrm{p}}$},  at confidence level p $\in(0,1)$,  is defined as

\vspace{-.5cm}
{\small \begin{eqnarray}
\mathrm{ ES}_{\mathrm{p}} \equiv \frac{1}{1-\mathrm{p}}\int_{\mathrm{p}}^{1}q_u(F_L)du,\nonumber
 \end{eqnarray}}

\vspace{-.3cm}
\noindent where $q_u(\cdot)$ is the quantile function defined as $q_u(F_L)\equiv\inf\{\ell \in \R : F_L(\ell)\geq u\}$.
 \end{defn}

The risk measures  \ES{}  and  \VaR{}  are related by the expression

\vspace{-.5cm}
{\small \begin{eqnarray*}
\mathrm{ ES}_{\mathrm{p}} \equiv \frac{1}{1-\mathrm{p}}\int_{\mathrm{p}}^{1}\mathrm{VaR}_u du
 \end{eqnarray*}}

\vspace{-.3cm}
\noindent and it can be shown  that, if  $L$ is integrable, with continuous distribution function {\small$F_L(\cdot)$}, then {\small$\mathrm{ES}_{\mathrm{p}} =   \E(L|L\geq \mathrm{VaR}_{\mathrm{p}})$} (see McNeil et al., 2005).

\begin{remark}
In the literature one can find variations for the risk measure \ES, given in the Definition \ref{es}, such as \emph{tail conditional expectation} (TCE), \emph{worst conditional expectation} (WCE) and  \emph{conditional} VaR (CVaR). Besides having slightly different definitions,  all these risk measures are equivalent when the distribution function is continuous.
\end{remark}

In practice, in order to calculate the  \VaR{} and \ES{} values, one needs to estimate the loss distribution function {\small$F_L(\cdot)$}. Obviously, the use of different methods to estimate  {\small$F_L(\cdot)$} will lead to different values of those measures. The most common approaches to calculate the \VaR{} are:

\begin{enumerate}
\item {\bf Empirical VaR}. This is a non-parametric approach. The empirical $\mathrm{VaR}_{\mathrm{p}}$  is the p-quantile of the empirical distribution function of the data. Under this approach, the  $\mathrm{VaR}_{\mathrm{p}}$ of $h$ periods is the same as the  $\mathrm{VaR}_{\mathrm{p}}$ of 1 period.

\item {\bf Normal VaR or Variance-Covariance Method}. Under this approach we assume that the data is normally distributed with mean and variance constants. The $\mathrm{VaR}_{\mathrm{p}}$ is then the p-quantile of the Gaussian distribution. The mean and the variance (or the covariance matrix if the data is multidimensional) are estimated by their sample counter parts. For this approach it is also very common to assume that the conditional distribution function of the data is Gaussian instead of the distribution function itself.

\item {\bf RiskMetrics Approach}. This methodology was developed by J.P. Morgan and it considers the conditional distribution function of the data. Consider first the case in which  the portfolio has only one asset. Let $r_t$ be the return (or log-return) of the portfolio at time $t$ (the loss is then $-r_t$). The methodology assumes that

\vspace{-.4cm}
{\small $$
 r_t|\F{t-1}\sim\mathcal{N}(\mu_t,\sigma_t^2),
 $$}

\vspace{-.5cm}
\noindent where the conditional mean and  variance  are such that

\vspace{-.4cm}
{\small\begin{equation}
 \mu_t=0\ \ \ \mbox{e}\ \ \ \sigma_t^2 = \lambda\sigma_{t-1}^2  + (1-\lambda)r_{t-1}^2,\ \ \ 0<\lambda<1,\label{sigma}
 \end{equation}}

 \vspace{-.5cm}
\noindent that is, $\{\sigma_t^2\}_{t\in\Z}$ follows an exponentially weighted moving average (EWMA) model (see Roberts, 1959).  Then, the $\mathrm{VaR}_{\mathrm{p}}$ at time $t+1$, is the p-quantile of the Gaussian distribution with mean $\mu_t$ and variance $\sigma^2_t$.

It can be shown that, using this method, the \VaR{} for a period $h$ is given by
{\small$\mathrm{ VaR}_{\mathrm{p},t}[h]=  \Phi^{-1}(\mathrm{p})\sqrt{h}\sigma_{t+1},$} where {\small$\Phi(\cdot)$} is the standard Gaussian distribution function and {\small$\sigma_{t+1}^2$} is defined by the expression \eqref{sigma}. It is easy to see that {\small$\mathrm{VaR}_{\mathrm{p},t}[h]=\sqrt{h}\mathrm{VaR}_{\mathrm{p},t+1}$}. However, if  {\small$\mu_t\neq 0$}, in expression \eqref{sigma}, this equality  no longer holds.

The multivariate case assumes that the  conditional distribution  function of the data is a multivariate normal one and {\small$\cov(r_{i,t+1},r_{j,t+1})=\gamma_{ij,t+1}$} is estimated by the expression

\vspace{-.4cm}
{\small$$
\gamma_{ij,t} = \lambda\gamma_{ij,t-1}+(1-\lambda)r_{i,t-1} r_{j,t-1},\ \ \ \mbox{for } 0<\lambda<1.
$$}

\vspace{-.6cm}
For more details see Zangari (1996).

\item {\bf Econometric Approach}. This approach is similar to the RiskMetrics one. However, in this case, a more general class of models is considered. Generally, the time series mean is modeled by a linear model, such as the ARMA model, and the volatility is estimated by using a heteroskedastic model such as the FIEGARCH model defined in Section \ref{fiegarchsection}.

\end{enumerate}

In the following proposition we present an expression for  \ES{} under the normality assumption.

\begin{prop}
 Let $L$ be the random variable which represents the portfolio loss. If  $L$  has Gaussian distribution function with mean   $\mu$ and variance  $\sigma^2$ then,

\vspace{-.2cm}
{\small $$
 \ES{} =\mu + \sigma\frac{\phi\big(\Phi^{-1}(\mathrm{p})\big)}{1-\mathrm{p}}, \quad \mbox{for all}\quad  \mathrm{p}\in(0,1),
 $$}

\vspace{-.4cm}
\noindent where $\phi(\cdot)$ and  $\Phi(\cdot)$ are, respectively, the  density  and the  distribution function of a standard normal random variable.
 \end{prop}

\proof
By setting {\small $Z = \dfrac{L - \mu}{\sigma}$ } and noticing that  {\small $\displaystyle\lim_{z\rightarrow\infty}\frac{1}{\sqrt{2\pi}}e^{-z^2/2} =0$}, the proof follows directly from the fact that {\small$\P(L\geq \mathrm{VaR}_{\mathrm{p}})=1-\mathrm{p}$} and  {\small$ \mathrm{ES}_{\mathrm{p}} = \phantom{\Big |} \E(L|L\geq \mathrm{VaR}_{\mathrm{p}})$}.
\qed


\subsection{A Scenario Based Risk Measure}
The maximum loss, denoted by MaxLoss, introduced by Studer (1997), can be viewed either as a risk measure or as  a systematic way of performing a stress test. This risk measure can be  viewed as the worst possible loss. In many cases, the worst scenario may not exist since the function to calculate the value of a portfolio may be unbounded from below. It is known that the probability of scenarios occurrence  which are very far away from the present market state is very low. Therefore, the idea is to restrict attention to scenarios under a certain admissibility domain, also denominated by \emph{confidence region}, that is, a certain set of scenarios with high probability of occurrence. For example, if we assume that the data has an elliptic distribution function, such as the t-Student or the Gaussian distribution, then the admissibility domain is an ellipsoid (see Studer, 1997).

\begin{defn}
 Given an admissibility domain  $A$, the \emph{maximum loss} of a portfolio contained in  $A$ is given by

\vspace{-.4cm}
{\small$$
\MaxLoss_A(f) \equiv  f(\nn{Z}_{AM}) - \min_{ \n{Z} \in A}\big\{f(\nn{Z})\big\},
$$}

\vspace{-.4cm}
\noindent where $f(\cdot)$ is the function that determines the portfolio's price and the  vector {\small $\nn{Z}_{AM}=(Z_{AM,1},\cdots Z_{AM,m})'$} represents the $m$ risk factors for the current market situation.
\end{defn}

The maximum loss is a coherent risk measure in the same sense as defined by Artzner et al. (1999). Furthermore, the maximum loss gives not only the loss dimension but also the scenario in which this loss occurs.

\begin{remark}
$\phantom{.}$
\begin{enumerate}
\item[(i)] Note that, in order to compute the maximum loss, we need to set a closed confidence region  $A$, with a certain probability p of occurrence. Then, an equivalent definition of maximum loss is the following

\vspace{-.35cm}
{\small$$
\MaxLoss_A(f) = \max\left\{f(\nn{Z}_{AM}) - f(\nn{Z}) : \nn{Z} \in A \ \mbox{and }\P(A) = \mathrm{p}\!\!\phantom{\Big|}\right\}.
$$}

\vspace{-.5cm}
\item[(ii)] Since $f(\cdot)$ gives  the  portfolio value, the expression  $f(\nn{Z}_{AM}) - f(\nn{Z})$ represents the loss $L$ (or $-L$ if $\nn Z$  is measured previously to $AM$) in the portfolio.
\end{enumerate}
\end{remark}

A portfolio is called \emph{linear} if the loss $L$ is a linear function with respect to each one of the risk-factors change. The following theorem gives the expression of the MaxLoss for a linear portfolio $\PP$ with risk-factors change normally distributed.

\begin{thm}\label{maxlossthm}
Let $\PP$ be a linear portfolio and $f(\cdot)$ be the function that determines the portfolio value. Then, $f(\nn X)=\nn a'\nn X$, where $\nn a \in \R^m$ is a vector of real constants and  $\nn X \in \R^m$, is the vector of risk-factors change. It follows that, given a confidence level  \emph{p}, the maximum loss of the portfolio is given by

\vspace{-.4cm}
{\small\begin{equation}
\MaxLoss = -\sqrt{c_\mathrm{p}}\sqrt{\nn a' \nn{\Sigma} \nn a },\label{maxloss}
\end{equation}}

\vspace{-.4cm}
\noindent where $ \nn{\Sigma}$ is the covariance matrix of the risk-factors change and $c_\mathrm{p}$ is the \emph{p}-quantile  $\mathcal{X}^2_m$ distribution function with  $m$ degrees of freedom. Moreover, the worst scenario is given by

\vspace{-.3cm}
{\small\begin{equation}
\nn Z^* = -\frac{\sqrt{c_\mathrm{p}}}{\sqrt{\nn a' \nn{\Sigma} \nn a}}\nn \Sigma\nn a.\nonumber
\end{equation}}
\end{thm}

\proof
See Theorem 3.15 in Studer (1997).
\qed

\begin{remark}
Studer (1997) considers the \emph{Profit and Loss distribution} (P\&L), which is the distribution of the random variable $-L$,  instead of the loss distribution. However, both analysis lead to similar results. The only difference is which tail of the distribution is being analyzed.  Considering  this fact, notice that expression \eqref{maxloss} is very similar to the expression for the Normal \VaR, which is $\sqrt{z_\mathrm{p}}\sqrt{\nn a' \nn{\Sigma} \nn a }$; the only difference lies in the scaling factor: $c_\mathrm{p}$ is the p-quantile of a $\mathcal{X}^2_m$ distribution with $m$ degrees of freedom, whereas $z_\mathrm{p}$ is the p-quantile of a standard normal distribution. Contrary to the \VaR, MaxLoss depends on the number of risk factors used in the model.
\end{remark}

%
%

\section{Simulation}\label{simulation}
\renewcommand{\theequation}{\thesection.\arabic{equation}}
\setcounter{equation}{0}
\renewcommand{\thetable}{\thesection.\arabic{table}}
\setcounter{table}{0}
\renewcommand{\thefigure}{\thesection.\arabic{figure}}
\setcounter{figure}{0}

In this section we present a simulation study related to the  estimation of the volatility of the  risk measures \VaR{} on FIEGARCH$(p,d,q)$ processes. A theoretical study related to the generation and the estimation of FIEGARCH$(p,d,q)$  processes, considering the same set of parameters used here, can be found in  Lopes and Prass (2009).

The simulation study considers five different models and the generated time series are the same ones used in Lopes and Prass (2009). The representation of the FIEGARCH process is the one proposed by Bollerslev and Mikkelsen (1996), given in Definition \ref{definitionfiegarch}, where {\small$Z_{i,t}\sim\mathcal{N}(0,1)$}, for {\small$i \in \{1,\cdots,5\}$}. For each model we consider $1000$ replications, with sample size {\small$n\in \{2000,5000\}$}. The value $n=2000$ was chosen since this is the approximated size of the observed time series considered in Section \ref{application} of this paper. The value  $n=5000$ was chosen to analyze the asymptotic properties for the estimators. In the following,  M$i$, for {\small $ i\in \{1,\cdots,5\}$}, denotes  the simulated FIEGARCH$(p,d,q)$  model, that is,
{\small \begin{center}
    \begin{tabular}{lll}
      M1: FIEGARCH(0,0.45,1) & M2: FIEGARCH(1,0.45,1) & M3: FIEGARCH(0,0.26,4) \\
    M4: FIEGARCH(0,0.42,1)    &  M5: FIEGARCH(0,0.34,1)  &\\
    \end{tabular}
\end{center}}
The parameters of the models used  in the simulation study, are given in Table \ref{tab1}. These values are similar to those found in the analysis of the observed time series (see Section \ref{application}).

\begin{table}[h]\vspace{-.5cm}
\centering\renewcommand{\arraystretch}{1.2}
\caption{Parameter Values for the Generated Models.}\label{tab1}	\vspace{.2cm}
\scriptsize{\begin{tabular}{|c|c|c|c|c|c|c|c|c|c|}
\hline \hline
  Model     & $\omega$  &   $ \beta_1$ &   $ \beta_2$ &   $ \beta_3$ &    $\beta_4$ & $\alpha_1$ &     $\theta$  &     $\gamma$ &         $ d$ \\
\hline\hline
 M1 &         0.00 &       0.45 &          - &          - &          - &          - &      -0.14 &       0.38 &       0.45 \\
\hline
  M2 &          0.00 &        0.90 &       - &          - &          - &        0.80 &       0.04 &       0.38 &       0.45 \\
\hline
 M3 &          0.00 &       0.22 &       0.18 &       0.47 &      -0.45 &          - &      -0.04 &        0.40 &       0.26 \\
\hline
 M4 &          0.00 &       0.58 &          - &          - &          - &          - &      -0.11 &       0.33 &       0.42 \\
\hline
   M5 &        0.00 &       0.71 &          - &          - &          - &          - &      -0.17 &       0.28 &       0.34 \\
\hline\hline
\end{tabular}}\vspace{-5cm}
\end{table}
\newpage

\subsection{Volatility Estimation}\label{volestimation}
In order to estimate the volatility  Lopes and Prass (2009) used the \emph{fgarch} function (from the S-Plus)  to fit the FIEGARCH$(p,d,q)$  models to the generated time series. The S-Plus code consider the expression  \eqref{fiegarch3} instead.  For each time series  only the $n-10$ first values are considered, where $n$ is the sample size. The remaining final $10$ values are used to estimate the forecast  error values. The  mean square  error ($mse$) is defined by {\small$mse =\frac{1}{re}\sum_{t=1}^{re}e^2_t,$} where $re=1000$ is the number of replications and {\small$e_t=\theta - \hat{\theta}$} represents the estimation error for the parameter $\theta$, where $\theta$ is any parameter given in Table \ref{tab1}. For each model, the final value was obtained from the expression {\small$\hat{\theta}= \frac{1}{re}\sum_{k=1}^{re} \hat{\theta}_k,$} where $\hat{\theta}_k$ is the $k$-th estimator for $\theta$ in the $k$-th replication, for {\small$k \in \{1,\cdots,re\}$}.

Lopes and Prass (2009) compare the  mean  of $1000$  generated values of   {\small$\sigma_{i,t+h}$} (a known parameter) and  {\small$X_{i,t+h}^2$},  with the mean of the  $h$-step  ahead  forecast values  {\small$\hat{\sigma}_{i,t+h}$}  and {\small$\hat{X}_{i,t+h}^2$}, by calculating the mean  square error ($mse$) values,  for {\small $t=n-10$, $h\in$ $ \{1,\cdots,10\}$},  {\small$n\in\{2000,  5000\}$} and  $i=1,\cdots,5$. The authors observed that the higher the sample size, the higher the mean square forecast error. For {\small$\sigma_t$} (the square root of the volatility) the mean square forecast error values vary from $0.0037$ to $0.3227$ and, for  {\small$X_t^2$}, they vary from $2.0725$ to $12.1379$.

\subsection{\VaR{} Estimation}
In the following we present  estimation results of the risk measure \VaR{} for the generated time series.

In order to calculate the conditional mean and the conditional variance we use the $n-10$ first values of the generated time series, where $n$ is the sample size.  For each one of the $1000$ replications we used the approaches described in Subsection \ref{subvar} to obtain the estimated \VaR. For the Econometric approach we consider  EGARCH$(p,q)$ models, with $p=1=q$ and the FIEGARCH$(p,d,q)$ fitted to the time series in Lopes and Prass (2009), considering the same values of $p$ and $q$ used to generate the time series.

\begin{remark}
Since  {\small$Z_{i,t}\sim\mathcal{N}(0,1)$}, for {\small$i \in \{1,\cdots,5\}$}, the true value of the $\mathrm{VaR}_{\mathrm{p},i,t+1}$, where $i$ stands for the model and $t+1$ for the period,  is given by
{\small\begin{equation}
\mathrm{ VaR}_{\mathrm{p},i,t+1} = \Phi^{-1}(\mathrm{p})\times \sigma_{i,t+1},\ \ i=1,\cdots,5,\label{truevar}
 \end{equation}}

\vspace{-.5cm}
\noindent  where $t=n-10$,  $\Phi(\cdot)$ is the standard normal distribution function and  $\sigma_{i,t+1}^2$ is the value of the conditional variance (volatility) generated by the model M$i$, for $i=1,\cdots,5$.
 \end{remark}

Table \ref{tab22} presents the true value, given by expression \eqref{truevar}, and the estimated values of the risk measure \VaR, for $\mathrm{p}\in\{0.95;0.99\}$. The values in this table are the mean taken over  1000 replications and $n\in\{2000, 5000\}$.  The values for $n = 5000$ appear in parenthesis and $mse$ represents the mean square error value. By comparing the different approaches, we observe that the mean of the \VaR{} estimated values  are very close from each other. In all cases, the mean of the estimated values is higher than the mean for the true value of this risk measure, either for $\mathrm{p}=0.95$ or  $\mathrm{p}=0.99$. Also, there is little difference between the mean square error values when the sample size vary from  $n = 2000$ to $n = 5000$. In most cases, the empirical approach leads to estimators with higher mean square error values. However, there is little difference among the values from the Econometric and Normal approaches. Generally, those two approaches present better results than the RiskMetrics approach.

Lopes and Prass (2009) reported that the parameter estimation for the FIEGARCH models show coefficients with high mean square error values. The results presented in subsection \ref{volestimation}, show a high mean square error value for the volatility estimation, which was expected since the volatility estimation depends on the  parameter estimation value. As a consequence, although the underlying process is a FIEGARCH process, the Econometric approach using this model do not present better results.

\begin{table}[htbp] \vspace{-.5cm}\renewcommand{\arraystretch}{1.3}
\centering \caption{Mean Estimated Values of the Risk Measure \VaR{} under Different  Approaches for  Sample Sizes $n =2000$ and, in Parenthesis, $n=5000$, with $\mathrm{p}\in\{0.95;0.99\}$.}\vspace{.2cm}\label{tab22}
{\scriptsize\begin{tabular}{|c|c|c|c|c|}
\hline\hline
Approach &  $\phantom{\Big|}\widehat{\mathrm{VaR}}_{0.95}$  &        $mse$ &  $\widehat{\mathrm{VaR}}_{0.99}$  &        $mse$ \\
 \hline\hline
\multicolumn{ 5}{|c|}{ $\phantom{\Big|}$ M1; $n = 2000 \ (n = 5000)$; $\mathrm{VaR}_{0.95} = 1.6726 \ (1.6673)$;   $\mathrm{VaR}_{0.99}= 2.3656 \ (2.3581)$ } \\
\hline
  Empirical  & 1.7996 (1.7967) & {\bf0.1378} ({\bf0.1287}) & 2.8355 (2.8577) & 0.5257 (0.5206) \\
\hline
    Normal   & 1.8458 (1.8433) & 0.1534 (0.1453) & 2.6108 (2.6072) & {\bf0.3051} ({\bf0.2895}) \\
\hline
Risk Metrics & 1.7159 (1.6912) & 0.2386 (0.2270) & 2.4370 (2.4029) & 0.4750 (0.4548)\\
\hline
    EGARCH   & 1.7707 (1.7533) & 0.1722 (0.1521) & 2.5044 (2.4797) & 0.3445 (0.3043)\\
\hline
  FIEGARCH   & 1.7759 (1.7567) & 0.1936 (0.1621) & 2.5117 (2.4845) & 0.3872 (0.3243)\\
\hline
\hline
\multicolumn{ 5}{|c|}{$\phantom{\Big|}$ M2; $n = 2000 \ (n = 5000)$; $\mathrm{VaR}_{0.95} = 1.6765 \ (1.6691)$;   $\mathrm{VaR}_{0.99}= 2.3710 \ (2.3607)$ }  \\
\hline
  Empirical  & 1.7773 (1.7689) & 0.1014 (0.1094) & 2.7020 (2.7229) & 0.3285 (0.3625) \\
\hline
    Normal   & 1.8004 (1.7938) & 0.1077 (0.1170) & 2.5462 (2.5369) & 0.2152 (0.2337) \\
\hline
Risk Metrics & 1.7287 (1.7192) & 0.1353 (0.1352) & 2.4425 (2.4291) & 0.2694 (0.2697) \\
\hline
    EGARCH   & 1.7494 (1.7342) & {\bf0.0925} ({\bf0.0902}) & 2.4743 (2.4527) & {\bf0.1850} ({\bf0.1805}) \\
\hline
  FIEGARCH   & 1.7499 (1.7439) & 0.1035 (0.1114) & 2.4749 (2.4664) & 0.2071 (0.2228) \\
  \hline
\hline
 \multicolumn{ 5}{|c|}{$\phantom{\Big|}$ M3; $n = 2000 \ (n = 5000)$; $\mathrm{VaR}_{0.95} = 1.6517 \ (1.6594)$;   $\mathrm{VaR}_{0.99}= 2.3360 \ (2.3469)$ }  \\
\hline
  Empirical  & 1.6852 (1.6968) & {\bf0.0326} ({\bf0.0326}) & 2.5002 (2.5208) & 0.0965 (0.0967) \\
\hline
    Normal   & 1.7028 (1.7129) & 0.0330 (0.0339) & 2.4086 (2.4224) & {\bf0.0654} ({\bf0.0674}) \\
\hline
Risk Metrics & 1.6864 (1.6805) & 0.0683 (0.0598) & 2.3858 (2.3778) & 0.1349 (0.1190) \\
\hline
    EGARCH   & 1.6886 (1.6976) & 0.0414 (0.0359) & 2.3882 (2.4009) & 0.0828 (0.0717) \\
\hline
  FIEGARCH   & 1.6895 (1.6950) & 0.0466 (0.0382) & 2.3894 (2.3973) & 0.0932 (0.0765) \\
 \hline
\hline
 \multicolumn{ 5}{|c|}{$\phantom{\Big|}$ M4; $n = 2000 \ (n = 5000)$; $\mathrm{VaR}_{0.95} = 1.6637 \ (1.6691)$;   $\mathrm{VaR}_{0.99}= 2.3360 \ (2.3607)$ }  \\
\hline
  Empirical  & 1.7643 (1.7967) & {\bf0.1194} ({\bf0.1197}) & 2.7593 (2.8391) & 0.4399 (0.4714) \\
\hline
    Normal   & 1.8064 (1.8383) & 0.1330 (0.1340) & 2.5549 (2.5998) & 0.4016 (0.2667) \\
\hline
Risk Metrics & 1.6987 (1.7078) & 0.1958 (0.1901) & 2.4115 (2.4250) & 0.3912 (0.3805) \\
\hline
    EGARCH   & 1.7478 (1.7610) & 0.1327 (0.1314) & 2.4719 (2.4906) & {\bf0.2654} ({\bf0.2629}) \\
\hline
  FIEGARCH   & 1.7546 (1.7656) & 0.1559 (0.1512) & 2.4816 (2.4971) & 0.3117 (0.3025) \\
\hline
\hline
 \multicolumn{ 5}{|c|}{$\phantom{\Big|}$ M5; $n = 2000 \ (n = 5000)$; $\mathrm{VaR}_{0.95} = 1.6763 \ (1.6484)$;   $\mathrm{VaR}_{0.99}= 2.3708 \ (2.3314)$ }  \\
\hline
  Empirical  & 1.8038 (1.8001) & {\bf0.1473} ({\bf0.1237}) & 2.9232 (2.8391) & 0.6476 (0.5871) \\
\hline
    Normal   & 1.8672 (1.8593) & 0.1708 (0.1464) & 2.6411 (2.6298) & {\bf0.3384} (0.2917) \\
\hline
Risk Metrics & 1.7402 (1.7078) & 0.2926 (0.2636) & 2.4790 (2.3665) & 0.5869 (0.5277) \\
\hline
    EGARCH   & 1.7863 (1.7200) & 0.1780 (0.1314) & 2.5264 (2.4906) & 0.3560 ({\bf0.2726}) \\
\hline
  FIEGARCH   & 1.7959 (1.7242) & 0.2088 (0.1557) & 2.5400 (2.4386) & 0.4176 (0.3114) \\
\hline
\hline
  \end{tabular}}
\end{table}

In the following we assume that the simulated time series represent log-returns and we denote by $\{r_{i,t}\}_{t = 1}^n$ the time series generated from the model M$i$, for $i=1,\cdots,5,$ where $n$  is the sample size.

In  practice, the volatility is not observable. Therefore, it is not possible to calculate the true value of the risk measure \VaR{} and the mean square error value for its estimation. Recall that {\small$\mathrm{VaR}_{\mathrm{p},i,t+1}$, $i=1,\cdots,5$}, represents the maximum loss that can occur with a given probability p in the instant $t+1$. Also, {\small$-r_{i,t+1}$, $i=1,\cdots,5,$}  can be understood as the loss at time $t+1$.
 The usual approach is then to compare   \VaR{} estimated values with the value of the observed log-returns (in our case, the simulated time series).

Table \ref{tab2} presents the mean value of $-r_{i,t+1}$ (which is known), for $i=1,\cdots,5,$ and the mean values of the estimated $\mathrm{VaR}_{\mathrm{p},i,t+1}$, for $t=n-10$, where $n$ is the sample size and $\mathrm{p}\in\{0.95;0.99\}$.
We observe that, for each model and each p value, different approaches lead to similar results. In most cases,  the value of the risk measure \VaR{} estimated under the RiskMetrics approach presents the smallest mean square error value. Also, the values estimated under this approach are closer to the observed values (the log-returns) than the ones estimated by the Empirical and Normal approaches. By comparing both, the RiskMetrics and Econometric approaches we observe almost no difference between the estimated values.
As in the previous case, we need to take into account that the estimation of the FIEGARCH models has strong influence in the results.

\begin{table}[htbp] \vspace{-.5cm}\renewcommand{\arraystretch}{1.2}
\centering \caption{Mean Square Error Values for the Risk Measure \VaR{} under Different  Approaches for  Sample Sizes $n =2000$ and, in Parenthesis, $n=5000$, with $\mathrm{p}\in\{0.95;0.99\}$.}\vspace{.2cm}\label{tab2}
{\scriptsize\begin{tabular}{|c|c|c|c|c|c|c|c|}
\hline\hline
\multicolumn{2}{|c|}{Approach}  &Empirical & Normal & RiskMetrics & EGARCH & FIEGARCH \\
\hline
\multirow{4}{*}{$r_{1.t+1} =$ 0.0649 (-0.0185)}& \multirow{2}{*}{ p = 0.95}
	&  4.5099  &  4.6781  & 4.4996  & {\bf3.5713}  & 3.6421\\
      &          & (4.7085) & (4.8815) & ({\bf4.6719}) & (4.8073) & (4.8570)\\
                                               & \multirow{2}{*}{ p = 0.99}
      &  4.5099  &  4.6781  & 4.4996  & {\bf3.5713}  & 3.6421\\
      &          & (4.7085) & (4.8815) & ({\bf4.6719}) & (4.8073) & (4.8570)\\
\hline
\hline
\multirow{4}{*}{$r_{2.t+1} =$ -0.0015 (0.0424)}& \multirow{2}{*}{ p = 0.95}
	&  4.4530  & 4.5403   &  4.5726  &  {\bf3.3268} & 3.3688\\
	&   	     & ({\bf4.2751}) & (4.3669) & (4.3423) & (4.3238) & (4.4094)\\	
                                               & \multirow{2}{*}{ p = 0.99}
      &  4.5099  &  4.6781  & 4.4996  & {\bf3.5713}  & 3.6421\\
      &          & (4.7085) & (4.8815) & ({\bf4.6719}) & (4.8073) & (4.8570)\\
\hline
\hline
\multirow{4}{*}{$r_{3.t+1} =$ -0.0380 (0.0353)}& \multirow{2}{*}{ p = 0.95}
	&   4.0809  &  4.1427  & 4.1591   &  {\bf2.9299} & 2.9587\\
	&           & ({\bf3.8915}) & (3.9452) & (3.9260) & (3.9708) & (3.9892)\\
                                               & \multirow{2}{*}{ p = 0.99}
      &  4.5099  &  4.6781  & 4.4996  & {\bf3.5713}  & 3.6421\\
      &          & (4.7085) & (4.8815) & ({\bf4.6719}) & (4.8073) & (4.8570)\\
\hline
\hline
\multirow{4}{*}{$r_{4.t+1} =$ -0.0369 (-0.0103)}& \multirow{2}{*}{ p = 0.95}
	&    4.6457  &  4.8021  &   4.7666 &  {\bf3.4045} &3.4854\\
	&  & ({\bf4.6180}) & (4.7760) & (4.7099) & (4.8067) & (4.8756)\\
                                               & \multirow{2}{*}{ p = 0.99}
      &  4.5099  &  4.6781  & 4.4996  & {\bf3.5713}  & 3.6421\\
      &          & (4.7085) & (4.8815) & ({\bf4.6719}) & (4.8073) & (4.8570)\\
\hline
\hline
\multirow{4}{*}{$r_{5.t+1} =$ -0.0589  (0.0525)}& \multirow{2}{*}{ p = 0.95}
	&   4.9339  &  5.1815  &  5.2978  &  {\bf3.6491} &3.7537\\
	&   & ({\bf4.3492}) & (4.5625) & (4.3588) & (4.3689) & (4.4497)\\
                                               & \multirow{2}{*}{ p = 0.99}
      &  4.5099  &  4.6781  & 4.4996  & {\bf3.5713}  & 3.6421\\
      &          & (4.7085) & (4.8815) & ({\bf4.6719}) & (4.8073) & (4.8570)\\
\hline
\hline

  \end{tabular}}

 \vspace{.2cm}
{\footnotesize  {\bf Note:} The mean square error value is measured with respect to the log-returns instead of  the true \VaR{}.}
\end{table}

%
%

\section{Analysis of Observed Time Series}\label{application}
\renewcommand{\theequation}{\thesection.\arabic{equation}}
\setcounter{equation}{0}
\renewcommand{\thetable}{\thesection.\arabic{table}}
\setcounter{table}{0}
\renewcommand{\thefigure}{\thesection.\arabic{figure}}
\setcounter{figure}{0}

In this section we present the estimation and the analysis of risk measures for a portfolio  \PP{} of stocks. This portfolio is composed by stocks of four Brazilian companies. These assets are denoted by  {\small $A_i$, $i=1,\cdots,4$,} where:

$A_1$: represents the  Bradesco stocks. \ \ $A_2$: represents the   Brasil Telecom stocks.

$A_3$: represents the  Gerdau stocks. \ \ \ \ $A_4$: represents the  Petrobr\'as stocks.
\vspace{.2cm}

These stocks are negotiated in the Brazilian stock market, that is, in the  S\~ao Paulo Stock, Mercantile \& Futures Exchange (Bovespa). The notation $A_M$ (or, equivalently, $A_5$) is used to denote the financial market. The  market portfolio values are represented by the S\~ao Paulo Stock Exchange Index (Bovespa Index or IBovespa).

Prass and Lopes (2009) present a comparison study on risk analysis  using CAPM model, \VaR{} and MaxLoss  on FIEGARCH processes. They consider the same portfolio \PP{}  considered here and they also calculate the vector $a = (a_1,a_2,a_3,a_4)$ of weights for this portfolio. The same weights found by Prass and Lopes (2009) are considered in this paper, that is, $(a_1,a_2,a_3,a_4) = (0.3381,\, 0.1813,\, 0.3087,\, 0.1719)$.

In the following we fixed:
\begin{itemize}

\item $c_i$ is the number of stocks of the asset $A_i$, $i = 1,\cdots,4$. It follows that
$ c_i=\dfrac{a_iV_0}{P_{i,0}}$, where $V_0$  is the initial capital invested in the portfolio,  $P_{i,0}$ is the unitary price of the stock at the initial time, and $\nn a = (a_1,a_2,a_3,a_4)'$ are the weights of the assets. The value of the portfolio \PP, at time $t$, is then given by
$$
V_t = V_0 \left(\frac{a_1}{P_{1,0}}P_{1,t}+\frac{a_2}{P_{2,0}}P_{2,t}+\frac{a_3}{P_{3,0}}P_{3,t}+\frac{a_4}{P_{4,0}}P_{4,t}\right),\ \ t= 1,\cdots,n,
$$
where $n$  is the sample size;

\item for this portfolio we assume that the initial time is the day of the first observation;

\item the risk-factors for this portfolio \PP{} are the logarithm of the prices of the assets. That is, the risk-factors vector is given by
$$
\nn Z_t = (Z_{1,t},Z_{2,t},Z_{3,t},Z_{4,t})'=(\ln(P_{1,t}),\ln(P_{2,t}),\ln(P_{3,t}),\ln(P_{4,t}))',
$$
where $P_{i,t}$ is the price of the asset  $A_i$,  $i=1,\cdots,4$, at time $t$. It follows that the risk-factors change is given by
$$
\nn X_t = (X_{1,t},X_{2,t},X_{3,t},X_{4,t})'=(r_{1,t},r_{2,t},r_{3,t},r_{4,t})',
$$
where $r_{i,t}$, is the log-return of the asset  $A_i$,  $i=1,\cdots,4$, at time $t$;

\item the loss of  the portfolio \PP, at time  $t$, is the random variable $L_t$  given by
\begin{eqnarray}
L_{t} = - V_{t-1} \sum_{i=1}^{4}a_i R_{i,t} \simeq - V_{t-1} \sum_{i=1}^{4}a_i r_{i,t} = -V_{t-1}r_{\mbox{\tiny{$\PP$}},t},\label{agoraprecisoperda}
\end{eqnarray}
where $V_{t-1}$ is the value of  \PP{}, at time  $t-1$, $\nn a =(a_1,a_2,a_3,a_4)'$ is the vector of weights, $R_{i,t} = \dfrac{P_t - P_{t-1}}{P_t}$,  $r_{i,t}$  is the log-return of the asset  $A_i$ at time $t$ and $r_{\mbox{\tiny{$\PP$}},t}$ is the log-return of the portfolio, at time $t$;

\item for \VaR{} and \ES{} estimation, in all cases (univariate or multivariate), we assume normality (or conditional normality) of the data.

\end{itemize}

\subsection{Characteristics of the Observed Time Series}

\begin{figure}[htbp]
\centering
\mbox{
\subfigure[]{\includegraphics[width=4.5cm,height= 3cm]{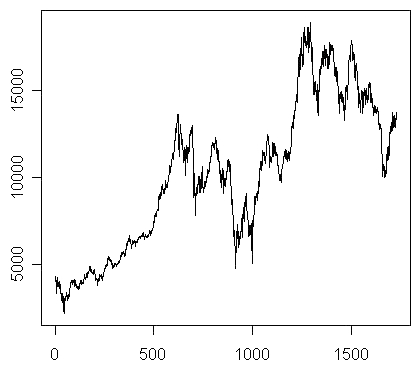}}\hspace{.1cm}
\subfigure[]{\includegraphics[width=4.5cm,height= 3cm]{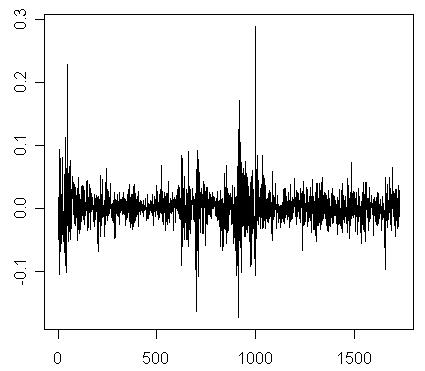}}\hspace{.1cm}
\subfigure[]{\includegraphics[width=4.5cm,height= 3cm]{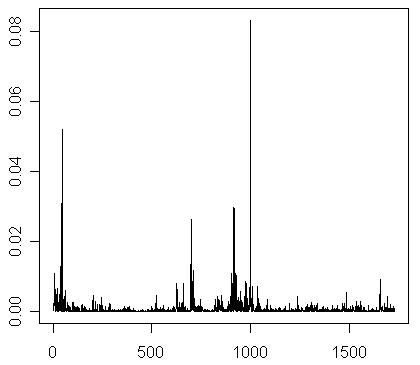}}
}
\caption{(a) S\~ao Paulo Stock Exchange Index (Bovespa Index or IBovespa) in the Period of January, 1995 to December, 2001;  (b) IBovespa Log-returns and (c) Square of the IBovespa Log-returns.}\label{fig1}
\end{figure}

\begin{figure}[h]
\centering
\mbox{
\subfigure[]{\includegraphics[height=3.1cm]{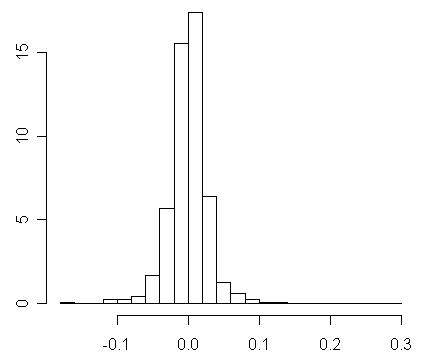}}\hspace{.1cm}
\subfigure[]{\includegraphics[height=3.1cm]{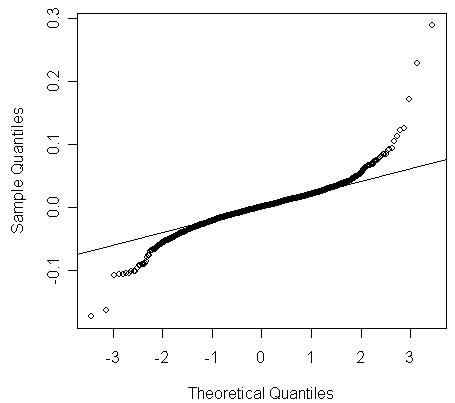}}\hspace{.1cm}
\subfigure[]{\includegraphics[height=3.1cm]{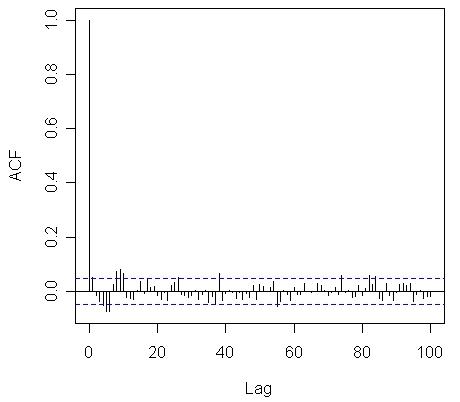}}\hspace{.1cm}
\subfigure[]{\includegraphics[height=3.1cm]{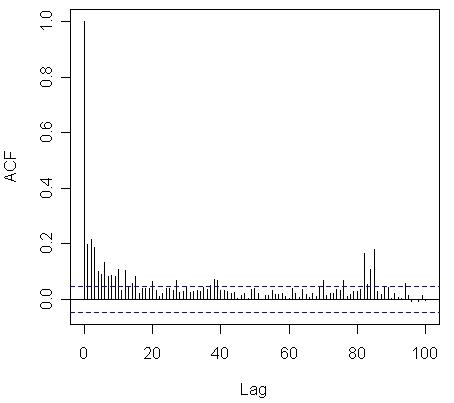}}
      }
\caption{(a) Histogram; (b) QQ-Plot and (c) Sample Autocorrelation of the  IBovespa Log-return series and (d) Sample Autocorrelation of the Square of the IBovespa Log-return series.}\label{fig2}
\end{figure}

Figure \ref{fig1} presents the time series with $n=1729$ observations of the S\~ao Paulo Stock Exchange Index (Bovespa Index or IBovespa) in the period of January, 1995 to December, 2001,  the  IBovespa log-return series and the square of the log-return series. Observe that the log-return series presents the stylized facts mentioned in Section \ref{fiegarchsection}, such as apparently stationarity, mean around zero and clusters of volatility. Also, in Figure \ref{fig2} we observe that, while the log-return series presents almost no  correlation, the sample correlation of the square of the log-return series assumes high values for several lags, pointing to the existence of both heteroskedasticity and long memory characteristics. Regarding the histogram and the  QQ-Plot, we observe that the distribution of the log-return series seems approximately symmetric and leptokurtic.

\begin{figure}[htbp]
\centering
\mbox{
\subfigure[Bradesco]{\includegraphics[width=6cm,height= 3cm]{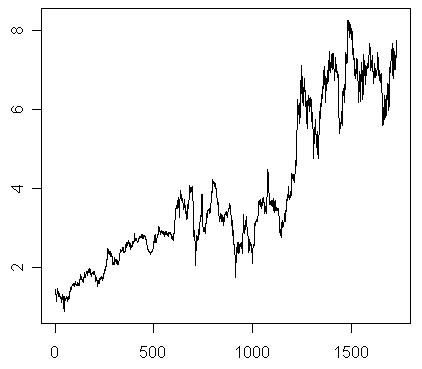}}\hspace{1cm}
\subfigure[Brasil Telecom]{\includegraphics[width=6cm,height= 3cm]{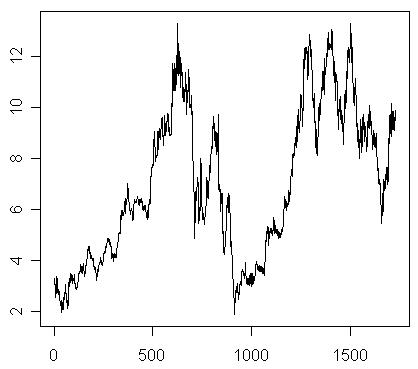}}
}
\mbox{
\subfigure[ Gerdau ]{\includegraphics[width=6cm,height= 3cm]{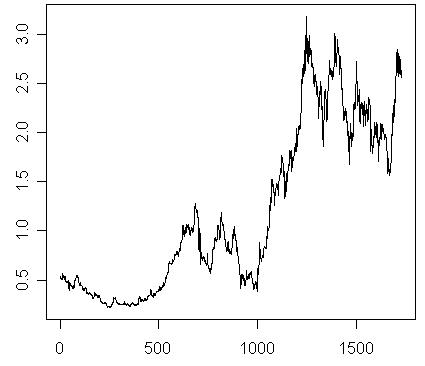}}\hspace{1cm}
\subfigure[Petrobr\'as]{\includegraphics[width=6cm,height= 3cm]{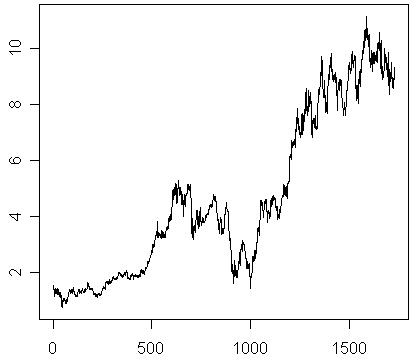}}
}
\caption{Time Series of the Stock Prices in the Period of January, 1995 to December, 2001. }\label{fig3}
\end{figure}

Figure \ref{fig3} presents the stock prices of the companies Bradesco, Brasil Telecom, Gerdau and Petrobr\'as,  in the period of January, 1995 to December, 2001, with $n=1729$ observations. These time series present the same characteristics observed in the IBovespa time series (and log-return time series).  In Figures \ref{fig1} and  \ref{fig3} we observe that both, the market index and the stock prices, present a strong decay in their values close to $t=1000$ (January 15, 1999). This period is characterized by the Real (the Brazilian currency) devaluation.

\begin{remark}
The presence of long memory was also tested by analyzing the periodogram of the square of the log-return time series  and by using some known hypothesis test such as GPH test, R/S, modified R/S, V/S and KPSS statistics. All these tests confirm the existence of long memory characteristics.
\end{remark}

\begin{figure}[h]
\centering
\mbox{
\subfigure[]{
  \includegraphics[width=4.5cm,height= 3cm]{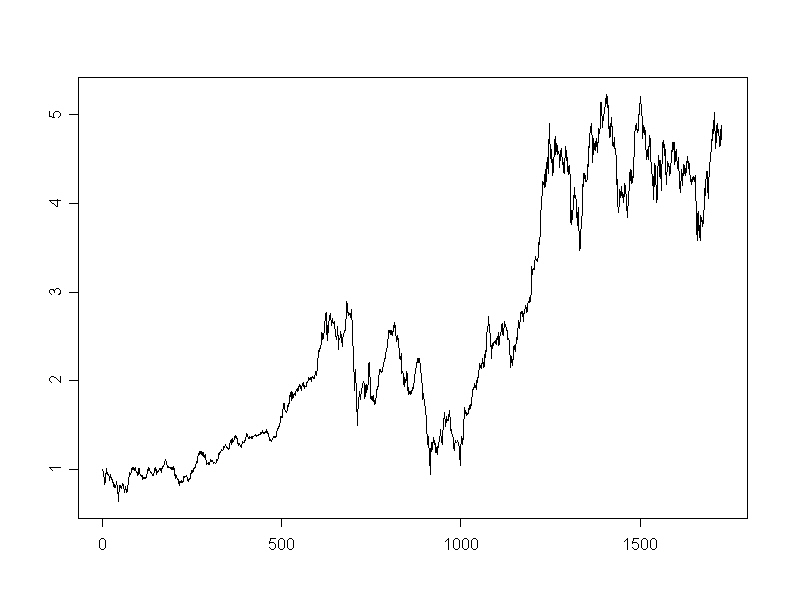}
 }\hspace{0.1cm}
  \subfigure[]{
  \includegraphics[width=4.5cm,height= 3cm]{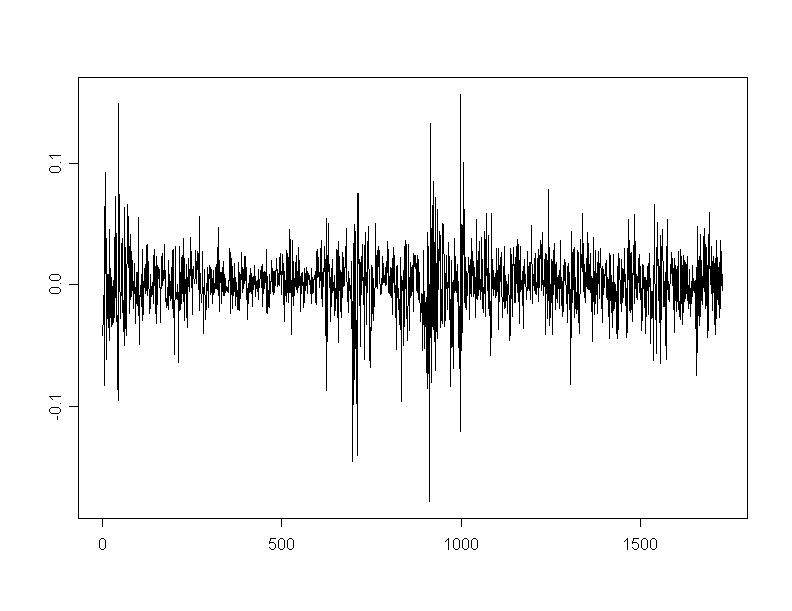}}\hspace{0.1cm}
  \subfigure[]{
  \includegraphics[width=4.5cm,height= 3cm]{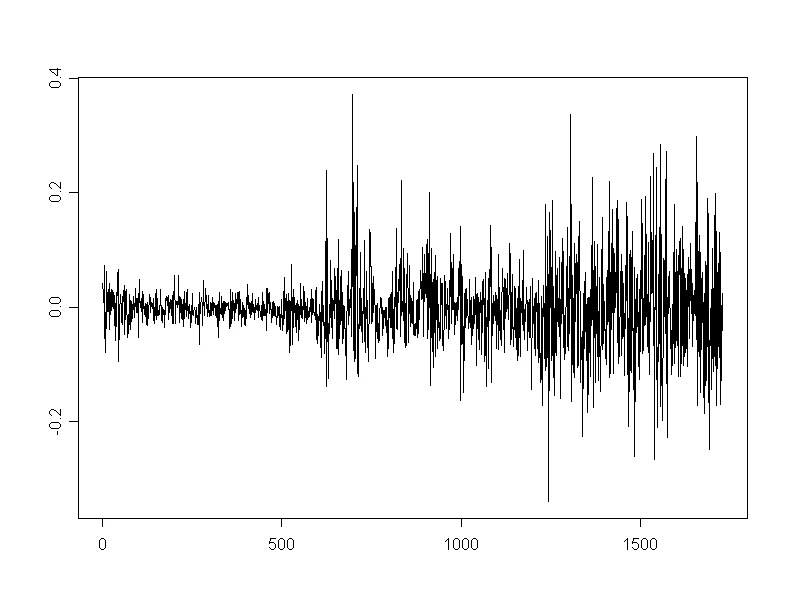}}
  }
  \caption{(a)  Time Series of the Portfolio $\PP$ Values;  (b)  Log-returns  and (c) Portfolio Loss for the Period of January, 1995 to  December, 2001. }\label{portivalue}
\end{figure}

Figure  \ref{portivalue}  presents the time series $\{V_t\}_{t=1}^{n}$ of the values of the portfolio \PP{} and the portfolio log-returns series in the period of January, 1995 to December, 2001. By comparing the portfolio values with the market index time series we observe a similar behaviour.

The time series of the portfolio loss is shown in  Figure  \ref{portivalue} (c). We observe that the highest loss occurred at  $t=698$ ,with $L_t=0.3729$  (this means that the  highest earning value is approximately equal to   R\$ 0.37 per Real invested) and the smallest loss occurred  at  $t=1244$, with $L_t=-0.3402$ (that is, approximately R\$ 0.34 per each Real invested). The highest loss corresponds to the change in the value of the portfolio \PP{} from  10/24/1997 (Friday) to 10/27/1997 (Monday). In this period the Bovespa index changed from  11,545.20 to 9,816.80 points, which represents a drop of  $14.97\%$. In this same date we also observed high drops in the  Dow Jones (7.18\%) and  S$\&$P 500  (6.87\%) indexes. This period coincides with the crises in Asia.  The highest earning in the portfolio \PP{} corresponds to the change in the value of the portfolio from  01/13/2001  to 01/14/2001. The IBovespa showed an increasing of  2.08\% in this period.   In the date  01/14/2000 the  Dow Jones index was  11,722.98 points. This value only was surpassed in   10/03/2006, when the index reached   11,727.34 points.

 \subsection{Fitting the FIEGARCH Models}
In the following we present the models fitted to the observed time series. The selection of the class of models used, the ARMA-FIEGARCH models, was based on the analysis of the sample autocorrelation and on  periodogram functions and results from the long memory tests. The fitted models are then used to estimate the volatility and consequently the risk measures for these processes.

In all cases considered, the analysis of the sample autocorrelation function  suggests an ARMA$(p_1,q_1)$-FIEGARCH$(p_2,d,q_2)$ model. The ARMA models are used to model the correlation among the log-returns while the FIEGARCH  models take into account the long memory and the heteroskedasticity characteristics of the time series.

In order to estimate the parameters of the model we consider the \emph{fgarch} function from S-PLUS software. We consider $p_1,q_1\in\{0,1,2,3\}$ e $p_2,q_2 \in\{0,1\}$. The selection of the final model was based on the values of the log-likelihood and on the  AIC and  BIC criteria. The residual analysis indicated that none of these models were adequated for the Gerdau log-return time series since the square of the log-returns still presented correlation. The problem was solved by choosing a FIEGARCH model with $p_2=4$. After the residual analysis, the following models were selected:

\begin{itemize}
\item For the IBovespa log-returns {\small$\{r_{M,t}\}_{t=1}^{1728}$:  ARMA$(0,1)$-FIEGARCH$(0,0.339,1)$},
{\scriptsize\begin{eqnarray*}
r_{M,t}& =&   X_{M,t} - 0.078X_{M,t-1}\\
X_{M,t} &=& \sigma_{M,t}Z_{M,t}\\
(1-0.706\LL)(1-\LL)^{0.339}\ln(\sigma_{M,t}^2) &=& -0.346 +0.275|Z_{M,t-1}| -0.166 Z_{M,t-1}.
\end{eqnarray*}}\vspace{-.6cm}

\item For the Bradesco log-returns {\small $\{r_{1,t}\}_{t=1}^{1728}$: ARMA$(0,1)$-FIEGARCH$(0,0.446,1)$},
{\scriptsize \begin{eqnarray}
r_{1,t}& =&   X_{1,t} - 0.129X_{1,t-1}\nonumber\\
X_{1,t} &=& \sigma_{1,t}Z_{1,t}\nonumber\\
(1-0.453\LL)(1-\LL)^{0.446}\ln(\sigma_{1,t}^2) &=&  -0.374 + 0.381|Z_{1,t-1}| -0.135 Z_{1,t-1}.\nonumber
\end{eqnarray}}\vspace{-.6cm}

\item For the Brasil Telecom log-returns {\small$\{r_{2,t}\}_{t=1}^{1728}$: ARMA$(0,1)$-FIEGARCH$(1,0.447,1)$},
{\scriptsize\begin{eqnarray}
r_{2,t}& =&  0.002+ X_{2,t} - 0.103 X_{2,t-1}\nonumber\\
X_{2,t} &=& \sigma_{2,t}Z_{2,t}\nonumber\\
(1-0.905\LL)(1-\LL)^{0.447}\ln(\sigma_{2,t}^2) &=&  -0.053 + 0.382|Z_{2,t-1}|-0.331 |Z_{2,t-2}|\nonumber\\
 &&+0.044 Z_{2,t-1} -0.066Z_{2,t-2}.\nonumber
\end{eqnarray}}\vspace{-.6cm}

\item For the Gerdau log-returns {\small$\{r_{3,t}\}_{t=1}^{1728}$: ARMA$(1,0)$-FIEGARCH$(0,0.256,4)$,}
{\scriptsize\begin{eqnarray}
r_{1,t}& =&  -0.1409r_{3,t-1}+ X_{3,t} \nonumber\\
X_{3,t} &=& \sigma_{3,t}Z_{3,t}\nonumber\\
\beta(\LL)(1-\LL)^{ 0.256}\ln(\sigma_{3,t}^2) &=&  -0.769 + 0.395|Z_{3,t-1}| -0.046 Z_{3,t-1},\nonumber
\end{eqnarray}}
where {\scriptsize$\beta(\LL)=1-0.216\LL- 0.184 \LL^2-0.470\LL^3+0.450\LL^4$.}

 \item For the Petrobr\'as log-returns {\small$\{r_{4,t}\}_{t=1}^{1728}$:  ARMA$(0,1)$-FIEGARCH$(0,0.416,1)$,}
{\scriptsize \begin{eqnarray}
r_{4,t}& =&  0.001 +  X_{4,t} -   0.119 X_{4,t-1}\nonumber\\
X_{4,t} &=& \sigma_{4,t}Z_{4,t}\nonumber\\
(1-0.575\LL)(1-\LL)^{0.416}\ln(\sigma_{4,t}^2) &=&   -0.347 + 0.326|Z_{4,t-1}| -0.110  Z_{4,t-1}.\nonumber
\end{eqnarray}}\vspace{-.6cm}

\item For the portfolio log-returns {\small $\{r_{\PP,t}\}_{t=1}^{1728}$: ARMA$(1,0)$-FIEGARCH$(0,0.233,1)$,}
{\scriptsize \begin{eqnarray}
r_{\mbox{\tiny{$\PP$}},t}& =&  -0.001 - 0.173r_{\mbox{\tiny{$\PP$}},t-1} + X_{\mbox{\tiny{$\PP$}},t}\nonumber\\
X_{\mbox{\tiny{$\PP$}},t} &=& \sigma_{\mbox{\tiny{$\PP$}},t}Z_{\mbox{\tiny{$\PP$}},t}\nonumber\\
(1- 0.754\LL)(1-\LL)^{0.233}\ln(\sigma_{\mbox{\tiny{$\PP$}},t}^2) &=& -0.498 +0.285|Z_{\mbox{\tiny{$\PP$}},t-1}| + 0.127 Z_{\mbox{\tiny{$\PP$}},t-1}.\nonumber
\end{eqnarray}}
 \end{itemize}

\subsection{Conditional Mean and Volatility Forecast}
In order to estimate the risk measures presented in Section \ref{riskmeasure} we first consider the estimation of the conditional mean and the volatility, that is, the  conditional standard deviation of the log-return time series. Theoretical results regarding forecast on ARMA and FIEGARCH models can be found, respectively, in
Brockwell and Davis (1991) and Lopes and Prass (2009).

The forecast for the conditional mean, $\hat{r}_{i,n+h}$, and for the volatility, $\hat{\sigma}_{i,n+h}$, for $i = 1, \cdots,5$ and $h = 1,\cdots,10$ are presented in Table \ref{tabforecast}. We observe that, for  $h>1$, the value of  $\hat\sigma_{M,n+h}$ is constant, which does not occur for the assets  $A_{i}$,  $i=1,\cdots,4$. Apparently, for $A_1$ and $A_4$ the forecast values increase, while for  $A_2$ they decrease. For  $A_3$ the behaviour seems to be random.

\begin{table}[h]
 \centering  \renewcommand{\arraystretch}{1.3}
\caption{Forecast Values of the Conditional Mean and Volatility  for the Log-return Time Series of IBovespa ($A_M$), Bradesco ($A_1$),  Brasil Telecom ($A_2$),  Gerdau ($A_3$) and   Petrobr\'as ($A_4$), for  $h$ = $1,\cdots,10$.}\vspace{.2cm}\label{tabforecast}
{\footnotesize\begin{tabular}{|c|c|c|c|c|c|c|c|c|c|c|}
\hline
\hline
& \multicolumn{ 2}{|c}{$A_M$} & \multicolumn{ 2}{|c}{$A_1$} & \multicolumn{ 2}{|c}{$A_2$} & \multicolumn{ 2}{|c}{$A_3$ } & \multicolumn{ 2}{|c|}{$A_4$} \\
\hline
$h$    &   $ \hat{r}_{M.n+h}$ & $\hat{\sigma}_{M.n+h}$ &    $\hat{r}_{1.n+h}$ & $\hat{\sigma}_{1.n+h}$&   $\hat{r}_{2.n+h}$ & $\hat{\sigma}_{2.n+h}$&   $ \hat{r}_{3.n+h}$ & $\hat{\sigma}_{3.n+h}$&   $\hat{r}_{4.n+h}$ & $\hat{\sigma}_{4.n+h}$ \\
\hline\hline
         1 &    -0.0012 &     0.0179 &     0.0031 &     0.0215 &    -0.0030 &     0.0319 &    -0.0041 &     0.0256 &     0.0019 &     0.0176 \\
\hline
         2 &     0.0000 &     0.0182 &     0.0000 &     0.0221 &     0.0018 &     0.0219 &     0.0000 &     0.0248 &     0.0014 &     0.0179 \\
\hline
         3 &     0.0000 &     0.0182 &     0.0000 &     0.0226 &     0.0018 &     0.0194 &     0.0000 &     0.0195 &     0.0014 &     0.0185 \\
\hline
         4 &     0.0000 &     0.0182 &     0.0000 &     0.0230 &     0.0018 &     0.0181 &     0.0000 &     0.0293 &     0.0014 &     0.0190 \\
\hline
         5 &     0.0000 &     0.0182 &     0.0000 &     0.0233 &     0.0018 &     0.0175 &     0.0000 &     0.0235 &     0.0014 &     0.0194 \\
\hline
         6 &     0.0000 &     0.0182 &     0.0000 &     0.0236 &     0.0018 &     0.0171 &     0.0000 &     0.0219 &     0.0014 &     0.0197 \\
\hline
         7 &     0.0000 &     0.0182 &     0.0000 &     0.0238 &     0.0018 &     0.0169 &     0.0000 &     0.0280 &     0.0014 &     0.0200 \\
\hline
         8 &     0.0000 &     0.0182 &     0.0000 &     0.0239 &     0.0018 &     0.0168 &     0.0000 &     0.0219 &     0.0014 &     0.0202 \\
\hline
         9 &     0.0000 &     0.0182 &     0.0000 &     0.0241 &     0.0018 &     0.0169 &     0.0000 &     0.0233 &     0.0014 &     0.0204 \\
\hline
        10 &     0.0000 &     0.0182 &     0.0000 &     0.0242 &     0.0018 &     0.0170 &     0.0000 &     0.0261 &     0.0014 &     0.0206 \\
\hline\hline
\end{tabular}}
\end{table}

The forecast for the conditional mean, $\hat{r}_{\PP,n+h}$, and for the volatility, $\hat{\sigma}_{\PP,n+h}$, of the portfolio \PP{} are presented in Table \ref{portiforecast}. Note that, while $\hat{r}_{\PP,n+h}$ is constant for $h>2$, $\hat{\sigma}_{\PP,n+h}$ is slowly increasing.

\begin{table}[htb]\vspace{-.5cm}
\centering  \renewcommand{\arraystretch}{1.2}
 \caption{Forecast Values of the Conditional Mean and Volatility  for the Portfolio Log-returns, for $h$ = $1,\cdots,10$.}\vspace{.2cm}\label{portiforecast}
{\footnotesize\begin{tabular}{|c|c|c|c|c|c|}
\hline\hline
        $ h$ &    $\hat{r}_{\mbox{\tiny{$\PP$}},t+h}$ & $\hat{\sigma}_{\mbox{\tiny{$\PP$}},t+h}$ & $ h$ &    $\hat{r}_{\mbox{\tiny{$\PP$}},t+h}$ & $\hat{\sigma}_{\mbox{\tiny{$\PP$}},t+h}$\\
\hline\hline
         1 &     0.0001 &     0.0149 & 6 &    -0.0010 &     0.0161 \\
\hline
         2 &    -0.0008 &     0.0152 & 7 &    -0.0010 &     0.0162 \\
\hline
         3 &    -0.0010 &     0.0155 & 8 &    -0.0010 &     0.0164 \\
\hline
         4 &    -0.0010 &     0.0157 & 9 &    -0.0010 &     0.0165 \\
\hline
         5 &    -0.0010 &     0.0159 & 10 &    -0.0010 &     0.0166 \\
\hline\hline
\end{tabular}}
\end{table}

\subsection{\VaR{} and \ES{} Estimation}\label{varEsEstimation}
We considered two different approachs in the analysis of the risk measures. The first one considers the log-return series of the portfolio \PP{} (see Palaro and Hotta, 2006) and its loss distribution.  We consider either, the conditional (RiskMetrics approach) and the unconditional (variance-covariance method) distribution of the risk-factors change in order to estimate the risk measures \VaR{} and \ES. As a second approach we calculate the risk measures for each one of the assets in the portfolio. Since \ES{} is a coherent risk measure and $\VaR{} \leq \ES{,}$ by calculating \ES{} we found an upper bound for the \VaR{} of the portfolio \PP.

Table  \ref{vares} presents the estimated values of \VaR{} and \ES{} for the portfolio log-return time series. The observed values, at time $n+1$, of the assets log-returns are, respectively, $-0.0026$, 0.0301, 0.0680 and 0.0021. Therefore, the portfolio log-return value, at this time, is 0.0259 (without loss of generality, we assume $V_n = 1$.  The loss is then given by a $L_{n+1}= -V_{n}\times r_{\mbox{\tiny{$\PP$}},n+1}=-0.0259$). By comparing this value with the estimated values given in the Table  \ref{vares} we observe that the Econometric approach, using FIEGARCH model, presents the best performance.

\begin{table}[h]\vspace{-.3cm}
\centering  \renewcommand{\arraystretch}{1.2}
\caption{Estimated Values of the Risk Measures \VaR{} and \ES{} for the Portfolio Log-return Time Series, at  Confidence Level  p = 95\% and, in Parenthesis, p = $99\%$, for $h=1$ day (Univariate Case).}\vspace{.2cm}\label{vares}
{\footnotesize \begin{tabular}{|c||c|c|}
\hline\hline
    $\phantom{\Big|}$ Approach &     $\mathrm{\VaR}_{\mathrm{p},n+1}$ &          $\mathrm{\ES}_{\mathrm{p},n+1}$ \\
\hline\hline
 Empirical &            0.0369 (0.0703) &           0.0588 (0.0966)\\
\hline
 Normal &       0.0398 (0.0566) &           0.0712 (0.0923) \\
\hline
 RiskMetrics &      0.0321 (0.0461) &            0.0583 (0.0759) \\
\hline
EGARCH &         0.0353 (0.0499) &          0.0625 (0.0808) \\
\hline
FIEGARCH &       0.0247 (0.0349) &       0.0437 (0.0564) \\
\hline\hline
\end{tabular}}
\end{table}

Table  \ref{varmulti} presents the results obtained by considering the multivariate distribution function of the risk-factor changes $\nn X_t = (r_{1,t},r_{2,t},r_{3,t},r_{4,t})' $. We consider the unconditional (Normal approach) and  the conditional distribution (RiskMetrics approach).  By comparing the results in Tables \ref{vares}  and \ref{varmulti} we observe that while for the Normal approach the values are the same, for the RiskMetrics approach the estimated value obtained using the univariate distribution was closer to the negative value of the observed log-return.

\begin{table}[h]\vspace{-.5cm}
\centering \renewcommand{\arraystretch}{1.2}
\caption{Estimated Values of the Risk Measures \VaR{} and \ES{} for the Portfolio Log-return Time Series, at  Confidence Level  p = 95\% and, in Parenthesis, p = $99\%$, for $h=1$ day (Multivariate Case).}\label{varmulti}\vspace{.2cm}
 {\footnotesize \begin{tabular}{|c|c|c|c|c|}
\hline\hline
     $\phantom{\Big|}$   Approach &     $\mathrm{\VaR}_{\mathrm{p},n+1}$ &      $\mathrm{\ES}_{\mathrm{p},n+1}$ \\
\hline\hline
 Normal &         0.0398 (0.0566) &        0.0712  (0.0923)\\
\hline
 RiskMetrics &          0.2131  (0.3018) &        0.3788 (0.4898) \\
\hline\hline
\end{tabular}}
\end{table}

Tables  \ref{varuniindi} and \ref{esuniindi} present the estimated values of the risk measures \VaR{} and \ES{} obtained by considering the univariate distribution function of each one of the risk-factor changes $r_{1,t},r_{2,t},r_{3,t},r_{4,t}$.  Upon comparison of the values in Tables \ref{varuniindi} and \ref{esuniindi} we observe that, for this portfolio \PP, both inequalities are satisfied:

{\small$$\vspace{-.1cm}
\mathrm{VaR}_{\mathrm{p},\mbox{\tiny{$\PP$}},n+1}\leq \sum_{i=1}^{4}a_i \mathrm{VaR}_{\mathrm{p},i,n+1}\ \ \ \mbox{and}\ \ \ \mathrm{ES}_{\mathrm{p},\mbox{\tiny{$\PP$}},n+1}\leq \sum_{i=1}^{4}a_i \mathrm{ES}_{\mathrm{p},i,n+1}.
$$}

\noindent Also, the \VaR{} estimated by Econometric approach using FIEGARCH processes were the ones closer to the observed loss  $-r_{i,1729}$,  $i=1,\cdots,4$, given by 0.0026,	 $-0.0301$,	 $-0.0680$	 and  $-0.0021$. It is easy to see that, in all cases, the loss was superestimated. This fact is well known and discussed in the literature. This occurs because of the normality assumption  in the risk measure estimation. Khindanova and Atakhanov (2002) presents a comparison study which demonstrate that stable modeling captures asymmetry and heavy-tails of returns, and, therefore, provides more accurate estimates of \VaR{}.

\begin{table}[h]\vspace{-.3cm}
\centering  \renewcommand{\arraystretch}{1.2}
\caption{\VaR{} Estimated Values for the Assets Log-return Time Series, at Confidence Level p = 95\% and, in Parenthesis, p = $99\%$, for $h=1$ day.}\label{varuniindi}\vspace{.2cm}
{\footnotesize\begin{tabular}{|c|c|c|c|c|c|}
\hline\hline
    Approach &        $\mathrm{VaR}_{\mathrm{p},1,n+1}$ & $\mathrm{VaR}_{\mathrm{p},2,n+1}$ & $\mathrm{VaR}_{\mathrm{p},3,n+1}$ & $\mathrm{VaR}_{\mathrm{p},4,n+1}$ & $\displaystyle\sum_{i=1}^{4}a_i\mathrm{VaR}_{\mathrm{p},i,n+1}$ \\
\hline\hline
   \multirow{2}{2.3cm} {\centering \small Empirical} &        0.0427 &     0.0508 &     0.0494 &     0.0483 &     0.0472 \\
           &          (0.0785) &     (0.1023) &     (0.0870) &     (0.0905) &     (0.0875) \\
\hline
   \multirow{2}{2.3cm} {\centering \small  Normal }&           0.0487 &     0.0582 &     0.0535 &     0.0539 &     0.0528 \\
           &            (0.0693) &     (0.0825) &     (0.0761) &     (0.0767) &     (0.0750) \\
\hline
  \multirow{2}{2.3cm} {\centering  \small RiskMetrics} &           0.2648 &     0.3056 &     0.3024 &     0.2511 &     0.2814 \\
           &          (0.3751) &     (0.4327) &     (0.4281) &     (0.3557) &     (0.3985) \\
\hline
  \multirow{2}{2.3cm} {\centering \small EGARCH} &          0.0370 &     0.0633 &     0.0565 &     0.0342 &     0.0473 \\
           &         (0.0536) &     (0.0875) &     (0.0783) &     (0.0489) &     (0.0666) \\
\hline
   \multirow{2}{2.3cm} {\centering\small FIEGARCH }&           0.0385 &     0.0495 &     0.0380 &     0.0308 &     0.0390 \\
           &           (0.0531) &     (0.0712) &     (0.0555) &     (0.0428) &     (0.0554) \\
\hline\hline
\end{tabular}}
\end{table}

\begin{table}[h]
\centering  \renewcommand{\arraystretch}{1.2}
\caption{\ES{} Estimated Values for the Assets Log-return Time Series, at Confidence Level p = 95\% and, in Parenthesis, p = $99\%$, for $h=1$ day.}\label{esuniindi}\vspace{.2cm}
{\footnotesize\begin{tabular}{|c|c|c|c|c|c|}
\hline\hline
   Approach &    $\mathrm{ES}_{\mathrm{p},1,n+1}$ & $\mathrm{ES}_{\mathrm{p},2,n+1}$ & $\mathrm{ES}_{\mathrm{p},3,n+1}$ & $\mathrm{ES}_{\mathrm{p},4,n+1}$ & $\displaystyle\sum_{i=1}^{4}a_i\mathrm{ES}_{\mathrm{p},i,n+1}$ \\
\hline\hline
 \multirow{2}{2.3cm} {\centering\small  Empirical} &         0.0672 &     0.0731 &     0.0623 &     0.0679 &     0.0669 \\
           &          (0.1133) &     (0.1163) &     (0.1049) &     (0.1202) &     (0.1124) \\
\hline
 \multirow{2}{2.3cm} {\centering\small Normal }&         0.0871 &     0.1037 &     0.0956 &     0.0964 &     0.0943 \\
           &         (0.1128) &     (0.1341) &     (0.1238) &     (0.1249) &     (0.1221) \\
\hline
 \multirow{2}{2.3cm} {\centering\small  RiskMetrics} &        0.4707 &     0.5428 &     0.5370 &     0.4464 &     0.5001 \\
           &         (0.6086) &     (0.7017) &     (0.6942) &     (0.5771) &     (0.6465) \\
\hline
  \multirow{2}{2.3cm} {\centering\small   EGARCH} &          0.0680 &     0.1085 &     0.0973 &     0.0616 &     0.0833 \\
           &         (0.0887) &     (0.1388) &     (0.1246) &     (0.0799) &     (0.1074) \\
\hline
 \multirow{2}{2.3cm} {\centering\small  FIEGARCH }&         0.0658 &     0.0901 &     0.0706 &     0.0532 &     0.0695 \\
           &        (0.0841) &     (0.1172) &     (0.0924) &     (0.0682) &     (0.0900) \\
\hline\hline
\end{tabular}}
\end{table}

\subsection{MaxLoss Estimation}

Since the considered portfolio \PP{} is linear, from Theorem  \ref{maxlossthm}, given a confidence level  p, the MaxLoss is estimated by the expression \eqref{maxloss}.

Table \ref{maxxlosssreais} presents the MaxLoss values under different values of p and the scenarios under which this loss occurs. By definition,  $r_{i,ML}$,  $i=1,\cdots,4$ is the log-return of the asset  $A_i$ under the MaxLoss scenario. By comparing the values in Table \ref{maxxlosssreais} with those ones found in the previous analysis, we observe that the loss estimated under the scenario analysis approach is higher than the loss estimated by \VaR{} and \ES{} (see Tables \ref{varuniindi} and \ref{esuniindi}). For all values of  p, the  MaxLoss value is higher (in absolute value) than the observed loss.

  \begin{table}[h]
 \centering \renewcommand{\arraystretch}{1.2}
 \caption{Portfolio Maximum Loss Values for Different Values of  p and Their Respective Scenario. }\label{maxxlosssreais}\vspace{.3cm}
\begin{tabular}{|c|c|c|c|c|c|}
\hline\hline
 & &    \multicolumn{ 4}{|c|}{Scenario} \\
\hline
         p & $MaxLoss$ &   $r_{1,ML}$ &   $r_{2,ML}$ &   $r_{3,ML}$ &   $ r_{4,ML}$ \\
\hline\hline
       0.50 &    -0.0453 &    -0.0451 &    -0.0460 &    -0.0453 &    -0.0449 \\
\hline
      0.55 &    -0.0475 &    -0.0473 &    -0.0482 &    -0.0475 &    -0.0471 \\
\hline
      0.65 &    -0.0521 &    -0.0519 &    -0.0529 &    -0.0521 &    -0.0517 \\
\hline
      0.75 &    -0.0574 &    -0.0572 &    -0.0583 &    -0.0574 &    -0.0569 \\
\hline
      0.85 &    -0.0642 &    -0.0640 &    -0.0652 &    -0.0642 &    -0.0637 \\
\hline
      0.95 &    -0.0762 &    -0.0759 &    -0.0773 &    -0.0762 &    -0.0756 \\
\hline
      0.99 &    -0.0901 &    -0.0897 &    -0.0915 &    -0.0901 &    -0.0894 \\
\hline\hline
\end{tabular}
\end{table}
\section{Conclusion}\label{conclusionsection}

Here we consider the same time series generated and analyzed in Lopes and Prass (2009) to estimate the risk measures \VaR, \ES{} and MaxLoss. For those time series we  fit FIEGARCH models to estimate the conditional variances of the time series. We observe that the higher the sample size, the higher the mean square error value. We use the estimated variances to estimate the risk measure \VaR{} for the simulated time series under different approaches. Since the estimated values for this risk measure, under different approaches, are very close from one another, we cannot say that one method is better than the others. For this simulated study, the Econometric approach, considering FIEGARCH models, does not perform as well as one would expected. However, the  results obtained by using these models have strong influence from the model parameter estimation, which is based on the quasi-likelihood method.  Asymptotic  properties of the quasi-likelihood estimator are still an open issue and it could explain the unexpected results.

Regarding the observed time series, we consider two different approaches for analyzing the  portfolio risk. We consider the distribution function of the portfolio log-returns (univariate case) and the multivariate distribution function of the risk-factor changes (multivariate case). Also, we consider either, the conditional and the unconditional distribution functions in all cases.

In the \VaR{} and \ES{} calculation, all approaches present similar results. By comparing  the observed loss, the values estimated using the econometric approach (and FIEGARCH models) were the closest to the observed values.  In all cases, the estimated loss was higher than the observed one. We also observe that the values estimated by considering the univariate distribution of the portfolio log-returns were smaller than the values estimated by considering the multivariate distribution of the risk-factor changes.  By comparing the estimated values of \VaR, \ES{} and MaxLoss  we observe that the loss estimated under the scenario analysis approach is higher than the loss estimated by \VaR{} and \ES. For all values of  p, the MaxLoss value is higher (in absolute value) than the observed loss.

\subsection*{Acknowledgments}

T.S. Prass was supported by CNPq-Brazil. S.R.C. Lopes research was
partially supported by CNPq-Brazil, by CAPES-Brazil, by {\it
INTC in Mathematics} and also by Pronex {\it Probabilidade e
Processos Estoc\'asticos} - E-26/170.008/2008 -APQ1.

\vspace{0.3cm}


\end{document}